\documentclass[twocolumn,showpacs,preprintnumbers,amsmath,amssymb]{revtex4}

\usepackage{graphicx}
\usepackage{dcolumn}
\usepackage{bm}
\usepackage{mathrsfs}
\usepackage{color}

\begin{document}

\title{General transformation of $\alpha$ cluster model wave function\\
to $jj$-coupling shell model 
in various $4N$ nuclei}%

\author{N. Itagaki$^1$, H. Matsuno$^2$, and T. Suhara$^3$}

\affiliation{
$^1$Yukawa Institute for Theoretical Physics, Kyoto University,
Kitashirakawa Oiwake-Cho, Kyoto 606-8502, Japan\\
$^2$Department of Physics, Kyoto University,
Kitashirakawa Oiwake-Cho, Kyoto 606-8502, Japan\\
$^3$Matsue College of Technology, Matsue, Shimane 690-8518, Japan\\
}

\date{\today}

\begin{abstract}
The antisymmetrized quasi-cluster model (AQCM) is a method 
to describe a transition from the $\alpha$-cluster wave function to the $jj$-coupling shell model
wave function. In this model, the cluster-shell transition is characterized by only two parameters; 
$R$ representing the distance between $\alpha$ clusters 
and $\Lambda$ describing the breaking of $\alpha$ clusters,
and the contribution of the spin-orbit interaction, very important in the 
$jj$-coupling shell model,
can be taken into account starting with the $\alpha$ cluster model wave function. 
In this article we show the generality of AQCM by extending 
the application to heavier region; various $4N$ nuclei from $^4$He to $^{52}$Fe.
We show and compare the energy curves for the $\alpha$+$^{40}$Ca cluster configuration
calculated with and without $\alpha$ breaking effect in $^{44}$Ti.
\end{abstract}

\maketitle

\section{Introduction}

General description of shell and cluster structures is a dream of nuclear structure physics.
Describing cluster states starting with shell models, including modern $ab\ initio$ ones, 
is a big challenge of computational nuclear science\cite{Maris,Dreyfuss,Yoshida}.
In general, we need huge model space in order to describe spatially correlating nucleons 
located at some place with some distance from the origin, 
when the single particle wave function is expanded with respect to the origin\cite{Haxton,Suzuki}.
One the other hand, if we start with the cluster model, we can easily describe cluster states
with much less computational efforts, but
the contribution of non-central interactions, very important in nuclear structure\cite{Mayer,Jensen}, 
vanishes
if we assume simple $(0s)^4$ configuration at some localized point for each $\alpha$ cluster.
The real systems have both natures of shell and cluster structures
and quantum mechanical mixing of these two plays a crucial role in many cases\cite{CSC,Suhara2015},
thus it is quite intriguing to establish a unified description
for the nuclear structure.
Our strategy is starting with the cluster model side,
contrary to the standard approaches starting with the shell model side,
and try to extend the model space to include
shell correlations especially for the contribution of 
the spin-orbit interaction.

As it is well-known, there is certain overlap between the shell and cluster model spaces.
When we take zero limit for the relative distance between clusters, the model space
coincides with the lowest shell model configuration owing to the antisymmetrization effect.
This is called Elliot SU(3) limit\cite{elli58}. For instance, if we put four $\alpha$ particles
in the form of tetrahedron shape and take the zero limit for the relative $\alpha$-$\alpha$ distances,
the wave function coincides with the closed shell configuration of the $p$-shell.
In this way, for $N = Z$ nuclei with magic numbers of three dimensional harmonic oscillator,
cluster model wave function can describe the doubly closed shell configurations
of major shells. 
Here both spin-orbit favored and unfavored single particle orbits are filled
and the contribution of the spin-orbit interaction cancels.
However, it is necessary to break $\alpha$ clusters to take into account the
spin-orbit contribution, if
only spin-orbit favored orbits of the last major shell are occupied.
This is crucial for nuclei corresponding the subclosure of the major shells.
In addition, the ``real" magic numbers of nuclear systems (beyond 20) 
appear indeed 
as a result of strong spin-orbit interaction,
which are different from the ones corresponding to the closed shell configurations
of three dimensional harmonic oscillator.

To overcome this difficulty of the cluster model,
we proposed the antisymmetrized quasi-cluster model (AQCM) \cite{Simple,Masui,Yoshida2,Ne-Mg,Suhara}, 
which enables us to describe the $jj$-coupling shell model states
with the spin-orbit contribution starting with the cluster model wave function.
In the AQCM, the transition from the cluster- to shell-model-structure 
can be described by only two parameters: $R$ representing the distance between $\alpha$ clusters, 
and $\Lambda$ which characterizes the transition of $\alpha$ cluster(s) to quasi-cluster(s) and quantifies the role of the spin-orbit 
interaction.
In Ref.~\cite{Suhara}, we have shown that the AQCM wave function in the limit of $R \rightarrow 0$ and 
$\Lambda = 1$ corresponds to the $(s_{1/2})^4 (p_{3/2})^8$ closed shell configuration of $^{12}$C,
and strong attractive contribution of the spin-orbit interaction can be taken into account. 
The optimal AQCM ground state of $^{12}$C is an intermediate state between the three-$\alpha$ cluster 
state and the shell model state with the $p_{3/2}$ subshell closure configuration. From a comparison with the 
antisymmetrized molecular dynamics (AMD) model, where all nucleons are treated independently, we found that
 the AQCM result is consistent with the AMD result (overlap is about 99\%) except for the small breaking
effect of threefold rotational symmetry.  
This result is quite surprising,
since the number of degrees of freedom in the AQCM trial wave function is significantly fewer than 
that in the AMD.

The purpose of the present work is to show the general applicability of the AQCM framework
in heavier region.
In this article, we apply AQCM to various $4N$ nuclei from $^4$He to $^{52}$Fe. 
We show that starting with the cluster model wave function, we can generate 
the $jj$-coupling shell model wave function with the spin-orbit contribution 
in these nuclei. 
We show and compare the energy curves for the $\alpha$+$^{40}$Ca cluster configuration
calculated with and without $\alpha$ breaking effect in $^{44}$Ti.


\section{The model}\label{model}

In this section, we describe the AQCM wave function in this work.

\subsection{Single particle wave function (Brink model)}

In conventional $\alpha$ cluster models, 
the single particle wave function has a Gaussian shape \cite{Brink};
\begin{equation}
	\phi_{i} = \left( \frac{2\nu}{\pi} \right)^{\frac{3}{4}}
	\exp \left[- \nu \left(\bm{r}_{i} - \bm{R}_i \right)^{2} \right] \eta_{i},
\label{Brink-wf}
\end{equation}
where $\eta_{i}$ represents the spin-isospin part of the wave function, 
and $\bm{R}_i$ is a real parameter representing the center of a Gaussian 
wave function for the $i$-th particle. 
In this Brink-Bloch wave function, four nucleons in one $\alpha$ cluster share the common $\bm{R}_i$ value. 
Hence, the contribution of the spin-orbit interaction vanishes. 

\subsection{Single particle wave function in the AQCM}

In the AQCM, $\alpha$ clusters are changed into quasi clusters. 
For nucleons in the quasi cluster, 
the single particle wave function is described by 
a Gaussian wave packet, and
the center of this packet $\bm{\zeta}_{i}$ is a complex parameter;
\begin{align}
	\psi_{i} &= \left( \frac{2\nu}{\pi} \right)^{\frac{3}{4}}
		\exp \left[- \nu \left(\bm{r}_{i} - \bm{\zeta}_{i} \right)^{2} \right] \chi_{i} \tau_{i}, 
\label{AQCM_sp} \\
	\bm{\zeta}_{i} &= \bm{R}_i + i \Lambda \bm{e}^{\text{spin}}_{i} \times \bm{R}_i, \label{center}
\end{align}
where
$\chi_{i}$ and $\tau_{i}$ in Eq.~\eqref{AQCM_sp} represent the spin and isospin part of the $i$-th 
single particle wave function, respectively. 
For the width parameter, we use the value of $b = 1.46$ fm, $\nu = 1/2b^2$.
The spin orientation is given by the parameter $\bm{\xi}_{i}$, 
which is in general a complex parameter, 
while the isospin part is fixed to 
be 'up' (proton) or 'down' (neutron),
\begin{align}
	\chi_{i} &= \xi_{i\uparrow} |\uparrow \ \rangle + \xi_{i\downarrow} |\downarrow \ \rangle,\\
	\tau_{i} &= |p \rangle \ \text{or} \ |n \rangle.
\end{align}
In Eq.~\eqref{center}, $\bm{e}^{\text{spin}}_{i}$ is a unit vector for the intrinsic-spin orientation, 
and $\Lambda$ is a real control parameter describing the dissolution of the $\alpha$ cluster. 
As one can see immediately, the $\Lambda = 0$ AQCM wave function, which has no imaginary part, 
is the same as the conventional Brink-Bloch wave function.
The AQCM wave function corresponds to the $jj$-coupling shell model wave function,
such as subshell closure configuration,
when $\Lambda = 1$ and $\bm{R}_i \rightarrow 0$.
The mathematical explanation for it is summarized in Ref.~\cite{Suhara}.

\subsection{AQCM wave function of the total system}

The wave function of the total system $\Psi$ is antisymmetrized product of these
single particle wave functions;
\begin{equation}
\Psi = {\cal A} \{ (\psi_1 \chi_1 \tau_1) (\psi_2 \chi_2 \tau_2) (\psi_3 \chi_3 \tau_3) \cdot \cdot \cdot \cdot (\psi_A \chi_A \tau_A)\}.
\label{total-wf}
\end{equation}  
The projection onto parity and angular momentum eigen states can be numerically performed 
when necessary.

\section{Results}\label{results}

In this section, we show how we transform the cluster wave function to 
the $jj$-coupling shell model one
in various nuclei from $^4$He to $^{52}$Fe.

\subsection{$^4$He and $^8$Be}

The nucleus $^4$He is a closed shell configuration of the lowest $s$ shell
in the shell model description, and this agrees with the simple
$\alpha$ cluster model wave function.

In $^8$Be, four additional nucleons occupy $p_{3/2}$ orbits 
in the $jj$-coupling shell model, and we describe it starting
with a wave function of two $\alpha$ clusters and transforming it.
Suppose that in the intrinsic frame, two $\alpha$ clusters are
set on the $x$ axis with the relative distance $R$.
This is realized by giving Gaussian center parameters 
($\bm{R}_i$ in Eq.~\eqref{Brink-wf}) in the following way;
$\bm{R}_1 = 
\bm{R}_2 =
\bm{R}_3 = 
\bm{R}_4 = -R{\bf e_x}/2$,
and
$\bm{R}_5 = 
\bm{R}_6 =
\bm{R}_7 = 
\bm{R}_8 = R{\bf e_x}/2$,
where $\bf e_x$ is a unit vector in the $x$ direction.
Here $\bm{R}_1$ and $\bm{R}_5$ are parameters for proton spin-up,
$\bm{R}_2$ and $\bm{R}_6$ are for proton spin-down,
$\bm{R}_3$ and $\bm{R}_7$ are for neutron spin-up,
and
$\bm{R}_4$ and $\bm{R}_8$ are for neutron spin-down nucleons.
Based on Eq.~\eqref{center}, we transform two $\alpha$ clusters
to quasi clusters.
Gaussian center parameters for the four protons are given in this way;
\begin{equation}
\bm{\zeta}_1 = -R(\bm{e_x} + i \Lambda \bm{e_y})/2,
\end{equation}
\begin{equation}
\bm{\zeta}_2 = -R(\bm{e_x} - i \Lambda \bm{e_y})/2,
\end{equation}
\begin{equation}
\bm{\zeta}_5 = R(\bm{e_x} + i \Lambda \bm{e_y})/2,
\end{equation}
\begin{equation}
\bm{\zeta}_6 = R(\bm{e_x} - i \Lambda \bm{e_y})/2,
\end{equation}
where ${\bf e_x}$ and ${\bf e_y}$ are unit vectors in the $x$ and $y$ direction,
respectively, 
and these are the same for the neutron part,
$\bm{\zeta}_3 = \bm{\zeta}_1$, 
$\bm{\zeta}_4 = \bm{\zeta}_2$,
$\bm{\zeta}_7 = \bm{\zeta}_5$,
and
$\bm{\zeta}_8 = \bm{\zeta}_6$.
When $\Lambda$ is set to zero, the wave function consisting of two quasi clusters
agrees with that of two $\alpha$ clusters.

\begin{table} 
 \caption{The expectation values of
 one-body spin-orbit operator $\sum_i {\bf l_i} \cdot {\bf s_i}$
(one-body $ls$) 
 and principal quantum number ($n$)
of $^8$Be together with $R$ (fm) and $\Lambda$ values.}
  \begin{tabular}{|c|c|c|c|} \hline 
   $R$ (fm)  &  $\Lambda$ & one-body $ls$ & $n$\\ \hline
   3.00   &  0.00  & 0.00 & 5.39\\ \hline 
   0.01   &  0.00  & 0.00 & 4.00 \\ \hline 
   0.01   &  0.50  & 1.60 & 4.00 \\ \hline 
   0.01   &  1.00  & 2.00 & 4.00 \\ \hline 
  \end{tabular}   \\ 
\label{obls_Be}
\end{table}

\begin{figure}[t]
	\centering
	\includegraphics[width=6.0cm]{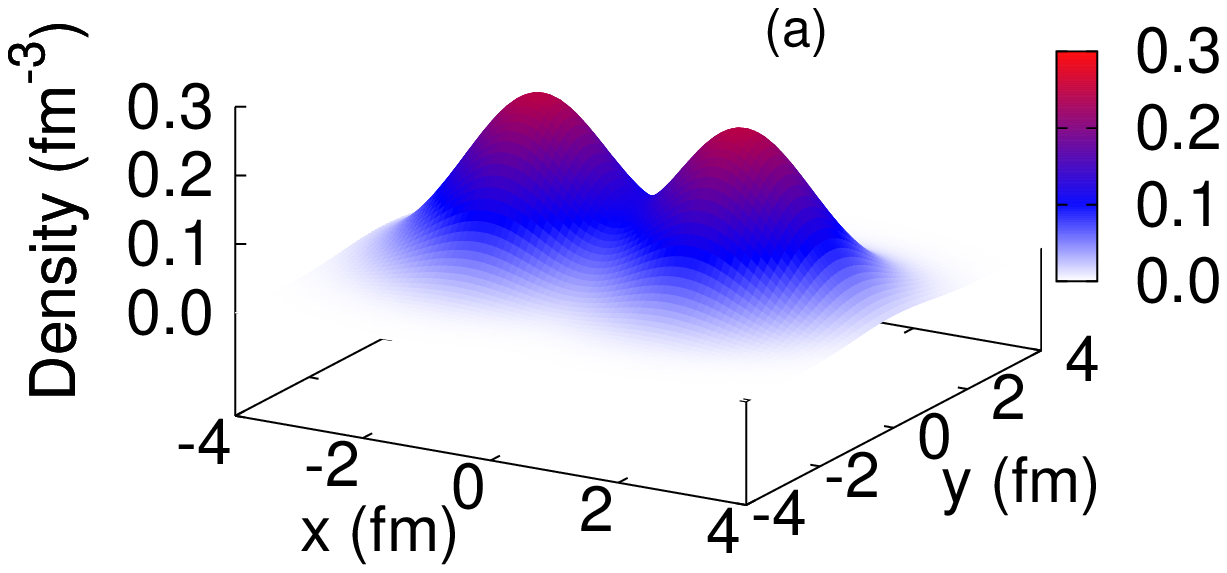} 
	\includegraphics[width=6.0cm]{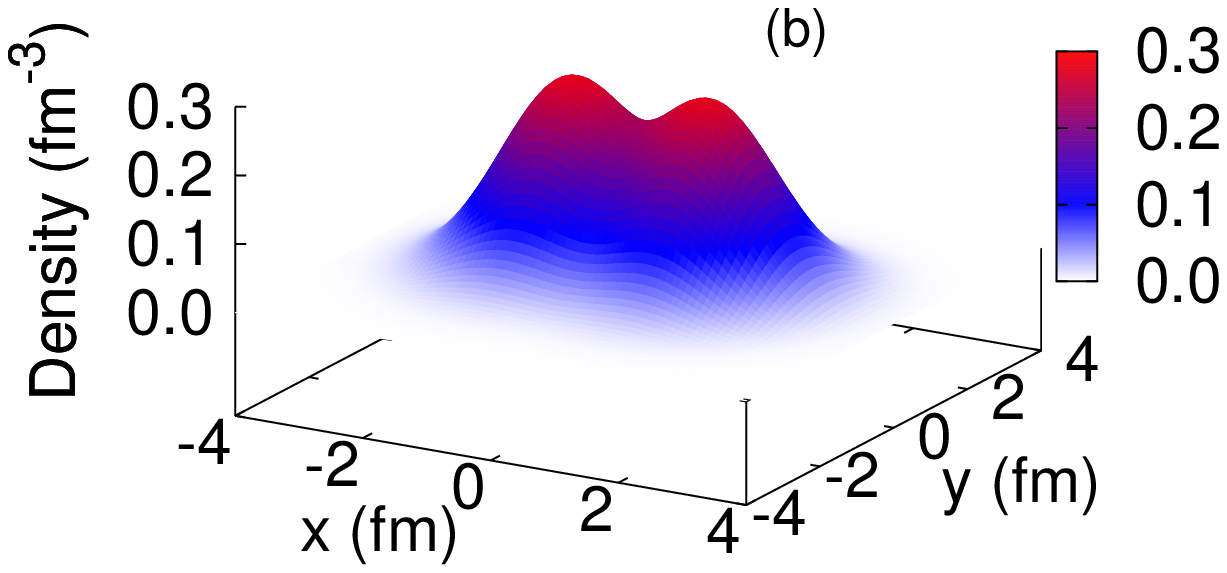}
	\includegraphics[width=6.0cm]{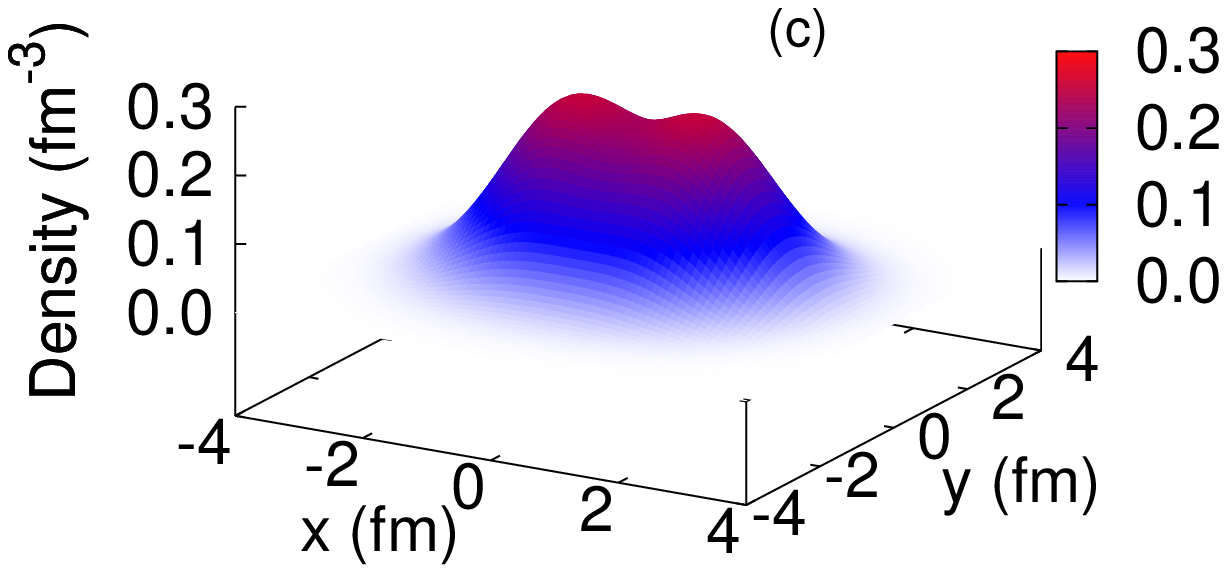}
    \includegraphics[width=6.0cm]{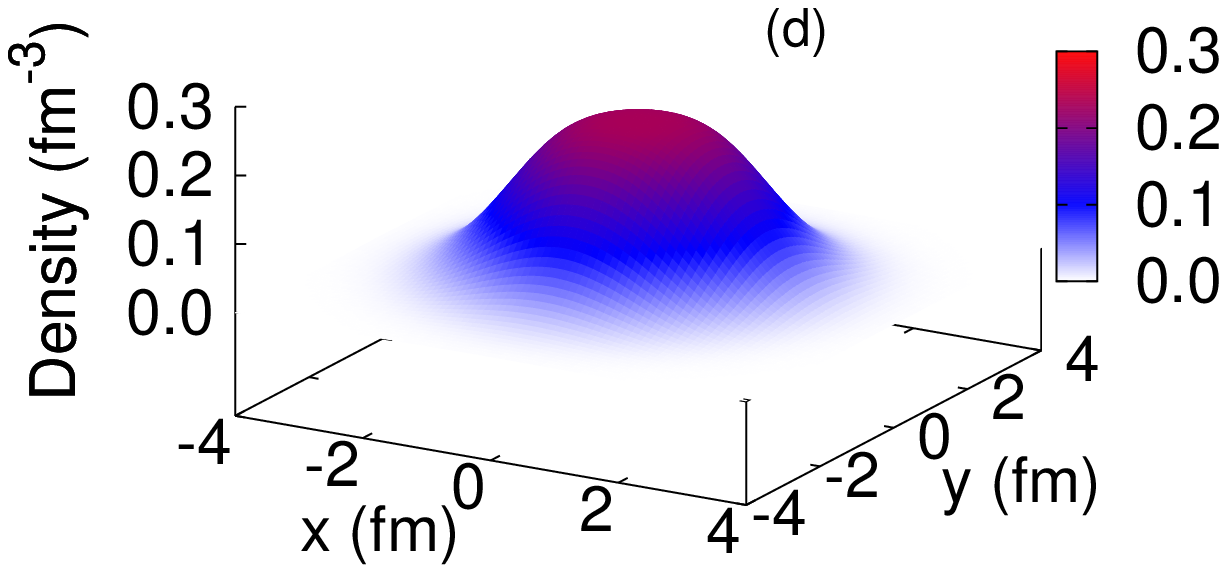} 
	\caption{(color online) The density plots of $^8$Be on $z=0$ plane 
    ($xy$ plane) as a function of $R$ and $\Lambda$.
    (a) $R = 3$ fm, $\Lambda =$ 0, 
    (b) $R = 0.01$ fm, $\Lambda =$ 0,
    (c) $R = 0.01$ fm, $\Lambda =$ 0.5, and
    (d) $R = 0.01$ fm, $\Lambda =$ 1. 
    The integrated density is normalized to the number of nucleons.
     }
	\label{8Be_density}
\end{figure}

The density distributions of $^8$Be as a function of $R$ and $\Lambda$
are shown in Fig.~\ref{8Be_density}, where
    (a) $R =
 3$ fm, $\Lambda =$ 0, 
    (b) $R = 0.01$ fm, $\Lambda =$ 0,
    (c) $R = 0.01$ fm, $\Lambda =$ 0.5, and
    (d) $R = 0.01$ fm, $\Lambda =$ 1.
In Fig.~\ref{8Be_density} (a), we can recognize two distinct 
peaks due to the $\alpha$ clusters. The relative distance $R$ is set to 3 fm,
and $\alpha$ clusters are not dissolved ($\Lambda = 0$).
These two peaks can be still clearly identified even if
we take the zero limit for the relative distance $R$
owing to the antisymmetrization effect (Fig.~\ref{8Be_density} (b) is the case
of $R$ = 0.01 fm). This is called Elliot SU(3) state and the first step
to transform cluster model wave function to the shell model one.
Then we change $\alpha$ clusters to quasi clusters by changing $\Lambda$
from zero to finite values; keeping $R = 0.01$ fm, the $\Lambda$ value is 
increased to 0.5 in (c) and 1.0 in (d). 
Figure~\ref{8Be_density} (c) is an intermediate state
between two-$\alpha$ cluster state and shell model state,
where $\alpha$ cluster structure partially remains but it starts melting.
In Fig.~\ref{8Be_density} (d)
we can see only one peak and $\alpha$ cluster structure is washed out.
Here the wave function is transformed to the $jj$-coupling shell model one.

The expectation values of
 one-body spin-orbit operator (one-body $ls$) 
 and principal quantum number 
of the harmonic oscillator ($n$)
of $^8$Be are listed in Table~\ref{obls_Be}
together with $R$ (fm) and $\Lambda$ values.
When $R = 3$ fm, the value of $n$ is 5.39, and this values becomes 4.00
at $R = 0.01$ fm,
which is the lowest possible value, where four of the
nucleons are excited from the $s$ shell to the $p$ shell.
The one-body spin-orbit operator ($\sum_i {\bf l_i} \cdot {\bf s_i}$)
is a good tool to see this transition, since the expectation value becomes zero
at $\Lambda = 0$,
and this value becomes 2.00 at $\Lambda =1$.
This is because the four nucleons in the $p_x$ shell are changed to
$p_{3/2}$ orbits of the $jj$-coupling shell model.
The eigen value for one nucleon 
in $p_{3/2}$ is \[\{j(j+1)-l(l+1)-s(s+1) \}/2 = \{ 15/4 -2 - 3/4\}/2 = 1/2\,\] 
and the value of 2.00 can be obtained by multiplying 4,
number of nucleons in this orbit.  

\subsection{$^{12}$C and $^{16}$O}
The transformation of 3$\alpha$ cluster wave function to 
subclosure configuration of $p_{3/2}$ shell in the $jj$-coupling shell model is 
discussed in Ref.\cite{Suhara} in detail.
The basic idea is the following; one $\alpha$ cluster 
comprised of spin-up proton, spin-down proton, spin-up neutron,
and spin-down neutron
is placed
on the $x$-axis and this is changed into quasi cluster as in
the $^8$Be case, and the second and third quasi clusters
are introduced by rotating both spatial and spin parts
of the first quasi cluster around the $y$-axis
by 120$^o$ and 240$^o$, respectively.

\begin{figure}[t]
	\centering
	\includegraphics[width=6.0cm]{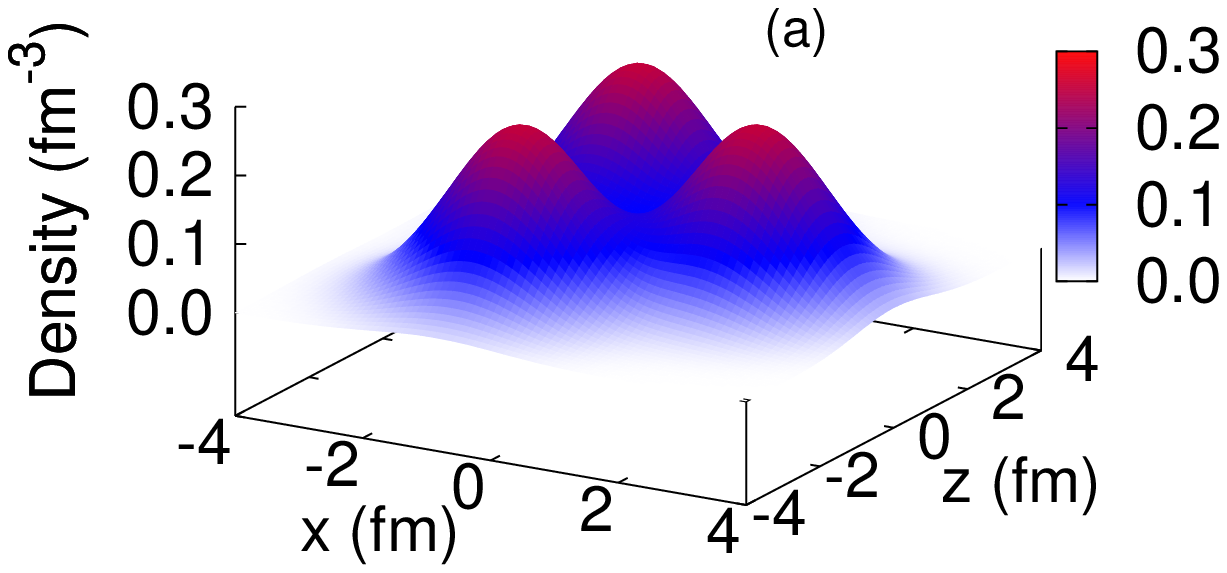} 
	\includegraphics[width=6.0cm]{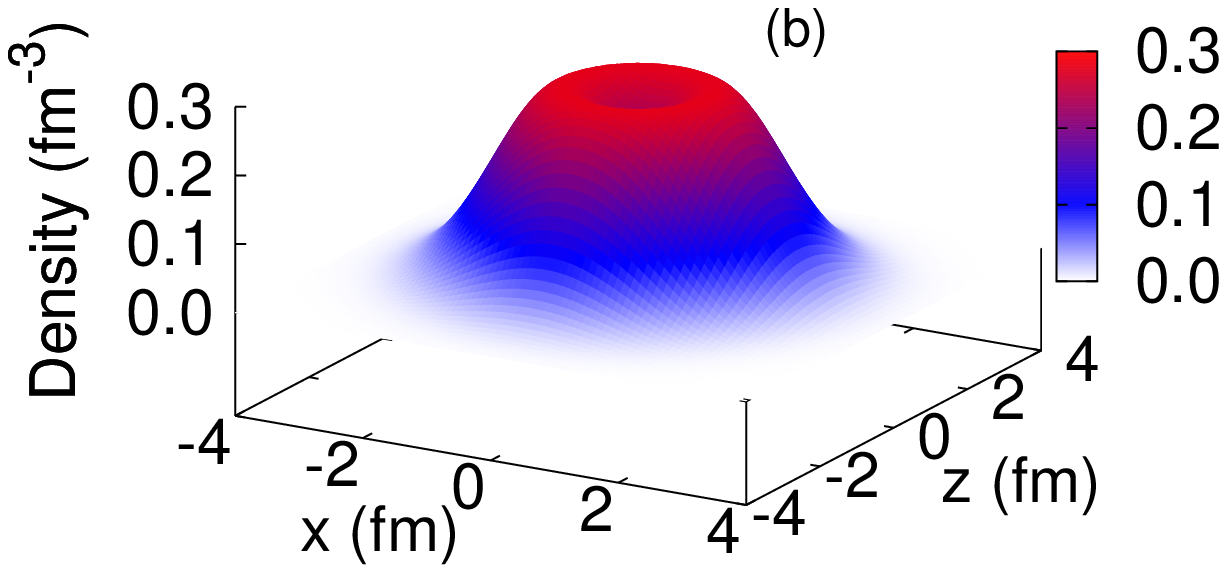}
	\includegraphics[width=6.0cm]{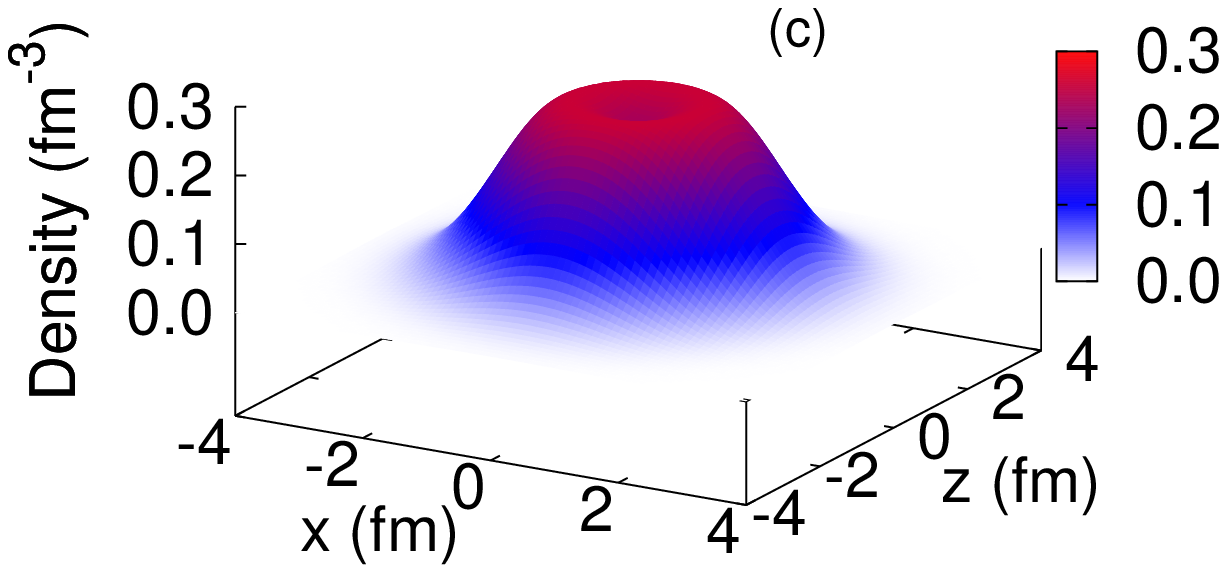}
    \includegraphics[width=6.0cm]{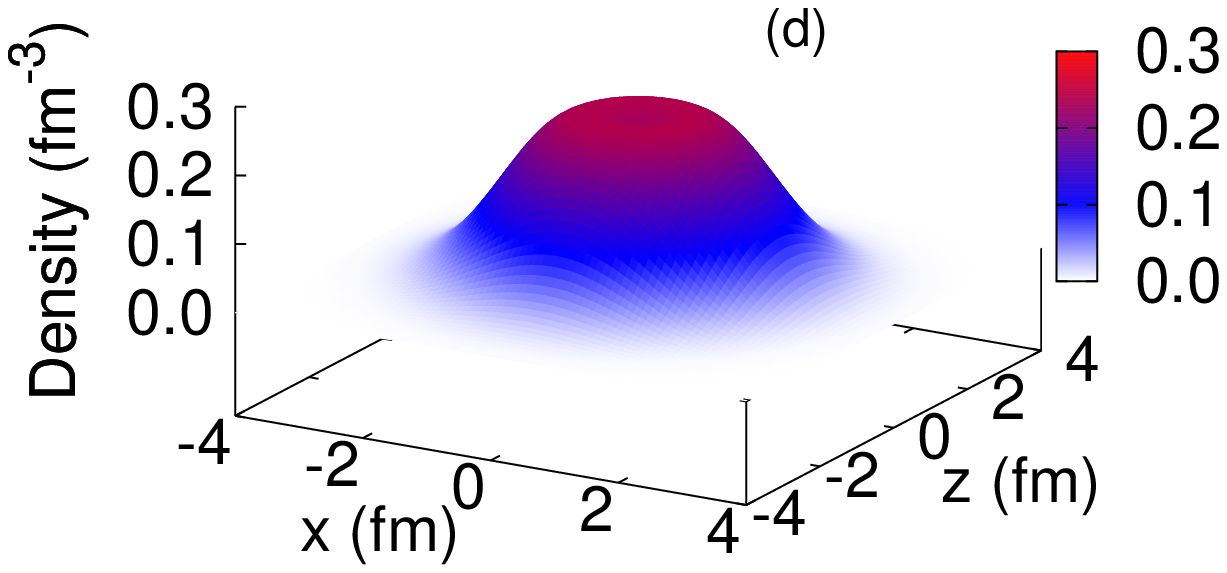} 
	\caption{(color online) The density plots of $^{12}$C on the $y=0$ plane ($xz$ plane) as a function of $R$ and $\Lambda$.
    (a) $R = 3$ fm, $\Lambda =$ 0, 
    (b) $R = 0.01$ fm, $\Lambda =$ 0,
    (c) $R = 0.01$ fm, $\Lambda =$ 0.5, and
    (d) $R = 0.01$ fm, $\Lambda =$ 1. 
    The integrated density is normalized to the number of nucleons.
     }
\label{12C_density}
\end{figure}

The density distributions of $^{12}$C on the $y=0$ plane ($xz$ plane)
as a function of $R$ 
(distance between quasi clusters) and $\Lambda$
are shown in Fig.~\ref{12C_density}, where
    (a) $R = 3$ fm, $\Lambda =$ 0, 
    (b) $R = 0.01$ fm, $\Lambda =$ 0,
    (c) $R = 0.01$ fm, $\Lambda =$ 0.5, and
    (d) $R = 0.01$ fm, $\Lambda =$ 1.
In Fig.~\ref{12C_density} (a), we can recognize three distinct 
peaks due to the $\alpha$ clusters. The relative distance $R$ is set to 3 fm
and $\Lambda = 0$.
These three peaks are changed into a donut shape if
we take the zero limit for the relative distance $R$
owing to the antisymmetrization effect (Fig.~\ref{12C_density} (b) is the case
of $R$ = 0.01 fm).
Then we change $\Lambda$
from zero to finite values; keeping $R = 0.01$ fm, $\Lambda$ value is 
increased to 0.5 in (c) and 1.0 in (d). 
We can see that middle part of the donut is
gradually filled and $\alpha$ cluster structure is washed out,
and the wave function is transformed to the $jj$-coupling shell model one.

\begin{figure}[t]
	\centering
	\includegraphics[width=6.0cm]{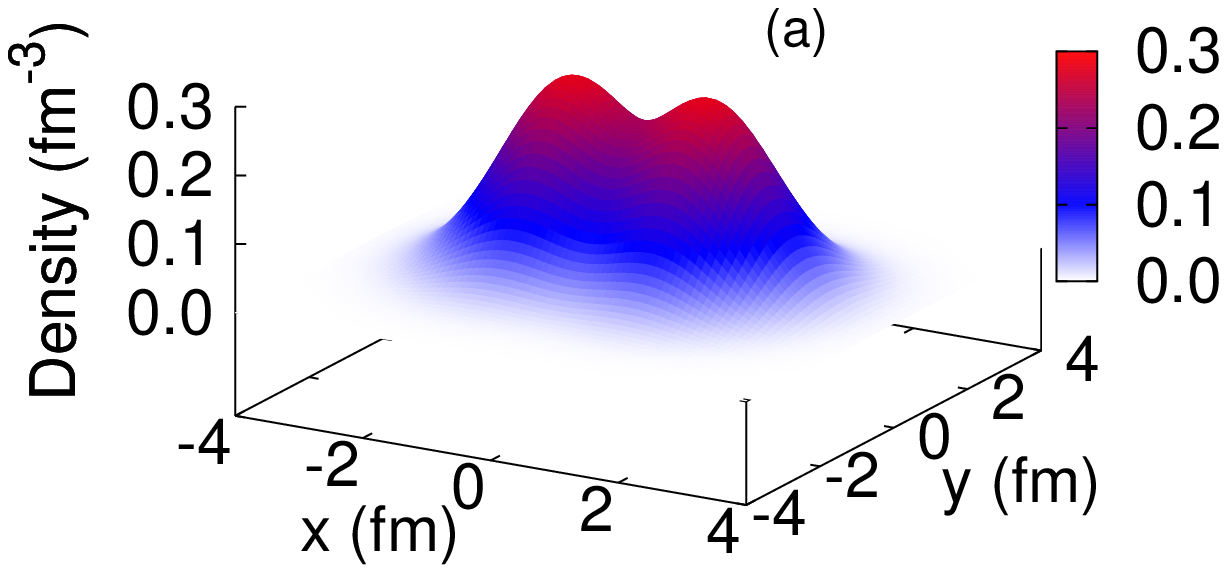}
    \includegraphics[width=6.0cm]{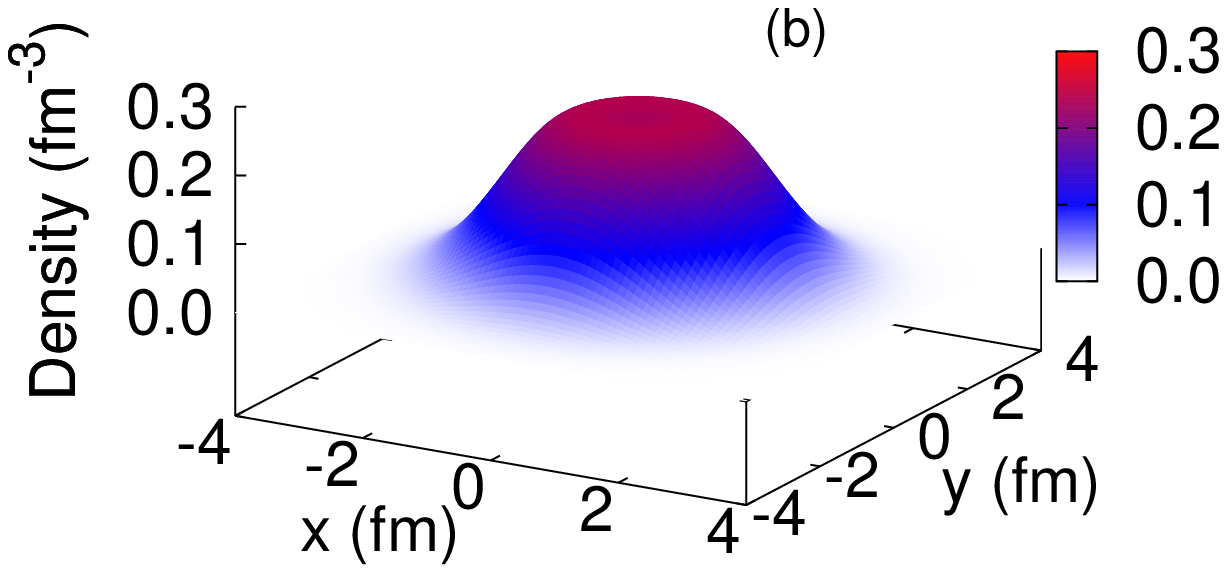} 
	\caption{(color online) The density plot of $^{12}$C on the $z=0$ plane ($xy$ plane) 
     as a function of $R$ and $\Lambda$.
    (a) $R = 0.01$ fm, $\Lambda =$ 0 and
    (b) $R = 0.01$ fm, $\Lambda =$ 1. 
    The integrated density is normalized to the number of nucleons.
     }
\label{12C_xy}
\end{figure}

It is quite instructive to compare 
the density distributions on the $y=0$ plane ($xz$ plane)
in Fig.~\ref{12C_density}
with the ones on the $z=0$ plane ($xy$ plane).
Figure~\ref{12C_xy} shows the density plot of $^{12}$C on the $z=0$ plane ($xy$ plane) 
with $R = 0.01$ fm and (a) $\Lambda =$ 0, and (b) $\Lambda =$ 1.
Here, Fig.~\ref{12C_density} (b) and Fig.~\ref{12C_xy} (a) are the same 
case of $R = 0.01$ fm and $\Lambda =$ 0, and
only the difference is the axes of the figures;
however they look quite different.
This means that even if inter cluster distances are very small ($R=0.01$ fm),
the nucleus is deformed and not spherical in the Brink model description for $^{12}$C.
On the other hand, if we introduce $\Lambda$ and transform $\alpha$ clusters 
to quasi clusters, we can describe spherical $^{12}$C
corresponding the subshell closure configuration;
Fig.~\ref{12C_density} (d) and Fig.~\ref{12C_xy} (b) are the same 
case of $R = 0.01$ fm and $\Lambda =$ 1, and the distributions 
on the $y=0$ plane ($xz$ plane) and that on the $z=0$ plane ($xy$ plane)
are very similar. The nucleus has the spherical symmetry.

\begin{table} 
 \caption{The expectation values of
 one-body spin-orbit operator $\sum_i {\bf l_i} \cdot {\bf s_i}$
(one-body $ls$) 
 and principal quantum number 
of the harmonic oscillator ($n$)
of $^{12}$Be together with $R$ (fm) and $\Lambda$ values.}
  \begin{tabular}{|c|c|c|c|} \hline
   $R$ (fm)  &  $\Lambda$ & one-body $ls$ & $n$\\ \hline
   3.00   &  0.00  & 0.00 & 11.22\\ \hline 
   0.01   &  0.00  & 0.00 & 8.00 \\ \hline 
   0.01   &  0.50  & 3.33 & 8.00 \\ \hline 
   0.01   &  1.00  & 4.00 & 8.00 \\ \hline 
  \end{tabular}   \\ 
\label{obls_C}
\end{table}

The expectation values of
 one-body spin-orbit operator
$\sum_i {\bf l_i} \cdot {\bf s_i}$ (one-body $ls$)
 and principal quantum number 
of the harmonic oscillator ($n$)
of $^{12}$C are listed in Table~\ref{obls_C}
together with $R$ (fm) and $\Lambda$ values.
When $R = 3$ fm, $n =$ 11.22, and this values becomes 8.00
at $R = 0.01$ fm,
which is the lowest possible value because eight
nucleons are excited to the $p$ shell.
The one-body $ls$ values are zero at $\Lambda =$ 0,
and this value becomes 4.00 at $\Lambda =1$,
interpreted as 0.5 (eigen value for a nucleon in $p_{3/2}$)
times 8 (number of nucleons in this orbit).  

For $^{16}$O, which corresponds to doubly closed shell of the shell model,
it is not necessary to introduce quasi clusters with $\Lambda$.
If we take the zero limit for the relative distances of $\alpha$ clusters
with tetrahedron configuration, the wave function corresponds to 
the doubly closed shell configuration of the $p$ shell
of the $jj$-coupling shell model,
where both $p_{3/2}$ and $p_{1/2}$ orbits are occupied.

\subsection{$^{20}$Ne and $^{24}$Mg}

The $jj$-coupling shell-model wave functions of
$^{20}$Ne and $^{24}$Mg are easily generated by adding
one or two quasi clusters around the $^{16}$O core,
which is introduced in the previous subsection.
Here the $^{16}$O core is described by a tetrahedron configuration 
of four $\alpha$ clusters, and the relative $\alpha$-$\alpha$ distances 
should be taken smaller than the value of parameter $R$, distance
between the quasi cluster(s) and the $^{16}$O core, to realize
the $jj$-coupling shell model wave function.
The detailed discussions are made in Ref.~\cite{Ne-Mg}.

\subsection{$^{28}$Si}

For $^{28}$Si,
as we briefly discussed in Ref.~\cite{Suhara},
there are two ways to transform $\alpha$ cluster wave function
to subclosure configuration of $d_{5/2}$ in the $jj$-coupling shell model.
One configuration is a pentagon shape of five quasi-$\alpha$ clusters 
on the $xz$-plane with the two $\alpha$ clusters placed on the $y$-axis.
Another configuration consists of a tetrahedron shape of the four $\alpha$ clusters 
whose center of gravity is at the origin and a triangle shape of 
three quasi-$\alpha$ clusters on the $xy$-plane surrounding four $\alpha$'s.
For the former case, one $\alpha$ cluster 
comprised of spin-up proton, spin-down proton, spin-up neutron,
and spin-down neutron
is placed
on the $x$-axis and this is changed into quasi cluster by
giving $\Lambda$ as in
the $^8$Be case, and the second, third, fourth, and fifth quasi clusters
are introduced by rotating both spatial and spin parts
of the first quasi cluster around the $y$-axis
by 72$^o$, 144$^o$, 216$^o$, and 288$^o$,
respectively. In addition, we place two $\alpha$ clusters on the $y$ axis. 
The latter is the combination of $^{12}$C and $^{16}$O wave functions
described in the previous subsection, and here twelve nucleons of $^{12}$C
are excited to the $sd$ shell due to the antisymmetrization effect.
If we take the zero limit for the relative distances among $\alpha$ clusters and quasi clusters,
these two configurations become identical and give the lowest shell model configuration of
$(s_{1/2})^4(p_{3/2})^8(p_{1/2})^4(d_{5/2})^{12}$ at $\Lambda = 1$. 

\begin{figure}[t]
	\centering
	\includegraphics[width=6.0cm]{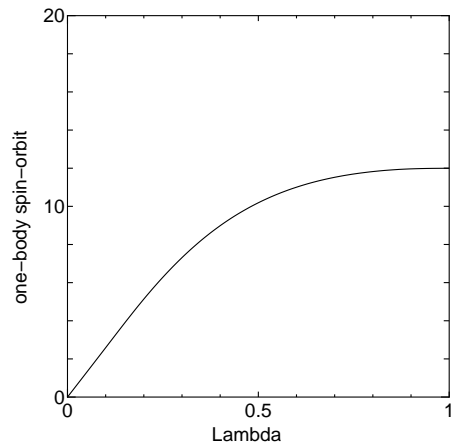} 
	\caption{The expectation value of the one-body spin-orbit operator 
     ($\sum_i {\bf l_i} \cdot {\bf s_i}$) for $^{28}$Si.
     The five quasi-$\alpha$ clusters form a pentagon shape
     on the $xz$-plane with the distance $R =$ 0.01 fm 
     and two $\alpha$ clusters are set on the $y$-axis.
	 The $\Lambda$ is changed from 0 to 1. 
     }
\label{Si-ls}
\end{figure}

\begin{figure}[t]
	\centering
	\includegraphics[width=6.0cm]{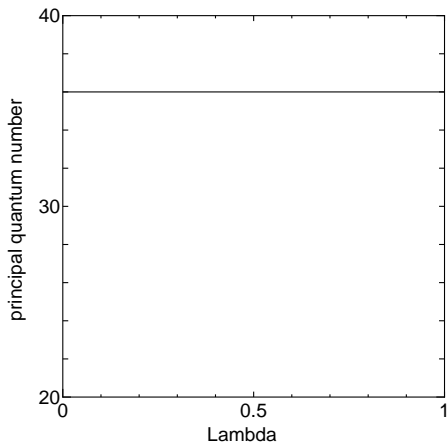} 
	\caption{The expectation value of the principal quantum number of the harmonic oscillator for $^{28}$Si.
     The five quasi-$\alpha$ clusters form a pentagon shape
     on the $xz$-plane with the distance $R =$ 0.01 fm 
     and two $\alpha$ clusters are set on the $y$-axis.
	 The $\Lambda$ is changed from 0 to 1. 
     }
\label{Si-n}
\end{figure}

The expectation value of the one-body spin-orbit operator 
$\sum_i {\bf l_i} \cdot {\bf s_i}$
for $^{28}$Si
is shown in Fig.~\ref{Si-ls}. The five quasi $\alpha$ clusters form 
a pentagon shape on the $xz$-plane with the distance $R =$ 0.01 fm, 
and $\Lambda$ is changed from 0 to 1. 
The eigen value for the one-body spin-orbit operator 
is 1 for a nucleon in the $d_{5/2}$ orbit,
and for $^{28}$Si, the value should be 12 because of twelve nucleons in $d_{5/2}$,
and this is achieved at $\Lambda$ = 1.
For this configuration, 
the expectation value of the principal quantum number of the harmonic oscillator
is shown in Fig.~\ref{Si-n}. The value is 36 because of twelve nucleons
in the $p$ shell and twelve nucleons in the $sd$ shell.
The value is constant and independent of the $\Lambda$ value.


\subsection{$^{32}$S and $^{36}$Ar}
For $^{32}$S and $^{36}$Ar, we have to transform 
the wave function of the nucleons in some of the $\alpha$ cluster(s)
to orbits where the spin-orbit interaction acts repulsively.
This can be performed by changing the sign of $\Lambda$ parameter for
some of the $\alpha$ clusters. 
The detailed analysis for $^{32}$S using AQCM is going on in Ref.~\cite{YYoshida}.

\subsection{$^{40}$Ca}

$^{40}$Ca corresponds to the doubly closed shell 
of the $sd$ shells, and cluster model and $jj$-coupling shell model give
the same representation 
without introducing quasi clusters with $\Lambda$.
Starting with the $\alpha$ cluster model, this is achieved by taking
the zero limit for the distances parameter $R$ of ten $\alpha$ clusters;
one $\alpha$ is centered at the origin, and other nine $\alpha$'s 
are at $R{\bf e_x}$, $-R{\bf e_x}$, $R{\bf e_y}$, $-R{\bf e_y}$,
$R{\bf e_z}$, $-R{\bf e_z}$, $R({\bf e_x}+{\bf e_y})/\sqrt{2}$,
$R(-{\bf e_y}+{\bf e_z})/\sqrt{2}$, and
$-R({\bf e_x}+{\bf e_z})/\sqrt{2}$,
where ${\bf e_x}$, ${\bf e_y}$, and ${\bf e_z}$
are unit vectors in the $x$, $y$, and $z$ direction, respectively. 
Here $R$ represents not the relative distances of $\alpha$ clusters
but the distance from the origin.
The Gaussian center parameters in Eq.~\eqref{Brink-wf} are given as follows;
\begin{equation}
\bm{R}_1 \sim \bm{R}_4 = 0,
\label{ca-1}
\end{equation}
\begin{equation}
\bm{R}_5 \sim \bm{R}_8 = R{\bf e_x},
\label{ca-2}
\end{equation}
\begin{equation}
\bm{R}_9 \sim \bm{R}_{12} = -R{\bf e_x},
\label{ca-3}
\end{equation}
\begin{equation}
\bm{R}_{13} \sim \bm{R}_{16} = R{\bf e_y},
\label{ca-4}
\end{equation}
\begin{equation}
\bm{R}_{17} \sim \bm{R}_{20} = -R{\bf e_y},
\label{ca-5}
\end{equation}
\begin{equation}
\bm{R}_{21} \sim \bm{R}_{24} = R{\bf e_z},
\label{ca-6}
\end{equation}
\begin{equation}
\bm{R}_{25} \sim \bm{R}_{28} = -R{\bf e_z},
\label{ca-7}
\end{equation}
\begin{equation}
\bm{R}_{29} \sim  \bm{R}_{32} = R({\bf e_x}+{\bf e_y})/\sqrt{2},
\label{ca-8}
\end{equation}
\begin{equation}
\bm{R}_{33} \sim \bm{R}_{36} = R(-{\bf e_y}+{\bf e_z})/\sqrt{2},
\label{ca-9}
\end{equation}
\begin{equation}
\bm{R}_{37} \sim \bm{R}_{40} = -R({\bf e_x}+{\bf e_z})/\sqrt{2}.
\label{ca-10}
\end{equation}

\begin{figure}[t]
	\centering
	\includegraphics[width=6.0cm]{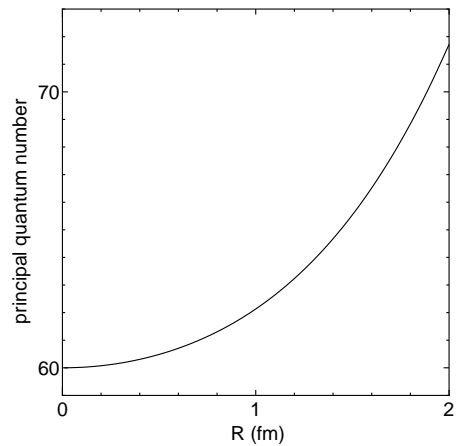} 
	\caption{The expectation value of the principal quantum number 
of harmonic oscillator for $^{40}$Ca as a function of the distance parameter $R$ in 
Eqs.~\eqref{ca-1}$\sim$\eqref{ca-10}.
     }
\label{Ca-n}
\end{figure}

The expectation value of the principal quantum number $n$ of harmonic oscillator 
for $^{40}$Ca is shown in Fig.~\ref{Ca-n} as a function of the distance parameter $R$ in 
Eqs.~\eqref{ca-1}$\sim$\eqref{ca-10}.
At $R = 0$ fm, the value converges to 60, reflecting the fact that
the twelve nucleons are in the $p$ shell and twenty four nucleons are
in the $sd$ shell. 

\subsection{$^{44}$Ti, $^{48}$Cr, and $^{52}$Fe}
For $^{44}$Ti,
the $^{40}$Ca+$\alpha$ cluster structure has been
widely discussed both in many theories and 
experiments\cite{Michel,Wada,Yamaya,Kimura}.
Also in $^{52}$Fe, possibility of 3$\alpha$ structure
around the $^{40}$Ca has been suggested\cite{Kokalova1,Kokalova2}.
The transition from such $\alpha$ cluster states
to the lowest $jj$-coupling shell model state
can be easily discussed within the present AQCM approach.

The $jj$-coupling shell-model wave functions of
$^{44}$Ti and $^{48}$Cr are easily generated by adding
one or two quasi clusters around the $^{40}$Ca core.
This is performed just by replacing the $^{16}$O 
core in our analysis for $^{20}$Ne and $^{24}$Mg in Ref.~\cite{Ne-Mg}
with the $^{40}$Ca core introduced in the previous subsection
($R$ in Eqs.~\eqref{ca-1}$\sim$\eqref{ca-10} is set to 0.01 fm).

\begin{figure}[t]
	\centering
	\includegraphics[width=6.0cm]{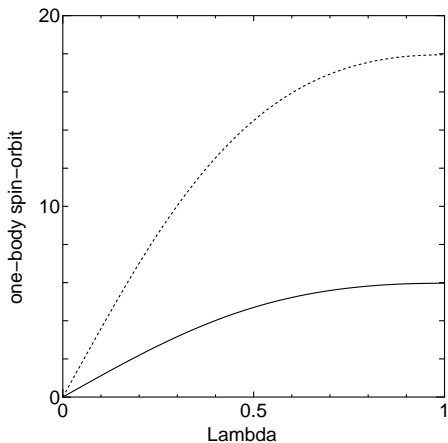} 
	\caption{The expectation value of the one-body spin-orbit operator 
     ($\sum_i {\bf l_i} \cdot {\bf s_i}$)
     for $^{44}$Ti (solid line).
	 One quasi cluster is placed on the $x$-axis with the distance of $R = 0.1$ fm
	 from the $^{40}$Ca core, and $\Lambda$ is changed from 0 to 1. The value for $^{52}$Fe is shown
	 in the dotted line, and distance parameter $R$ for the three quasi clusters is set to 0.25 fm.
	 For the $^{40}$Ca core part, $R$ in Eqs.~\eqref{ca-1}$\sim$\eqref{ca-10} is set to 0.01 fm,
     smaller than the value for the quasi cluster(s).
     }
\label{Ti-ls}
\end{figure}

The expectation value of the one-body spin-orbit operator 
($\sum_i {\bf l_i} \cdot {\bf s_i}$)
for $^{44}$Ti
is shown in Fig.~\ref{Ti-ls} (solid line).
One quasi cluster is placed on the $x$-axis with the distance of $R = 0.1$ fm
from the $^{40}$Ca core, and $\Lambda$ is changed from 0 to 1
($R$ in Eqs.~\eqref{ca-1}$\sim$\eqref{ca-10} for the $^{40}$Ca core part is set to 0.01 fm,
smaller than the value for the quasi cluster).
The eigen value for the one-body spin-orbit operator 
is 3/2
for a nucleon in the $f_{7/2}$ orbit, \[ \{j(j+1)-l(l+1)-s(s+1)\}/2=\{63/4-12-3/4\}/2=3/2, \]
and for $^{44}$Ti, the value should be 6 because of four nucleons in $f_{7/2}$.
In Fig.~\ref{Ti-ls} (solid line), we can recognize that this situation is achieved at $\Lambda$ = 1.

\begin{figure}[t]
	\centering
	\includegraphics[width=6.0cm]{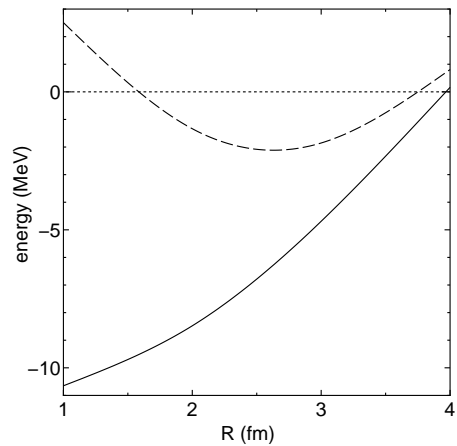} 
	\caption{The $0^+$ energy curves of $^{44}$Ti comprised of $^{40}$Ca and one quasi cluster
	as functions of relative distance $R$. 
    The dashed line is the case
	of $\Lambda$ = 0 for the quasi cluster ($^{40}$Ca+$\alpha$ model), 
    and for the solid line optimal value of $\Lambda$
	is chosen for each $R$ value. The energies are measured from
	the $^{40}$Ca+$\alpha$ threshold (dotted line).
     }
\label{Ca-alpha}
\end{figure}

The $0^+$ energy curves of $^{44}$Ti 
comprised of $^{40}$Ca and one quasi cluster
as functions of relative distance $R$ between them are shown in Fig.~\ref{Ca-alpha}.
When we calculate the expectation value of the Hamiltonian, we need larger value of
$R$ in Eqs.~\eqref{ca-1}$\sim$\eqref{ca-10}
for the $^{40}$Ca core to guarantee the numerical accuracy, 
and that is set to 0.1 fm.
The dashed line is the case
of $\Lambda$ = 0 for the quasi cluster, which is equivalent to an $\alpha$ cluster,
and for the solid line, the optimal value of $\Lambda$
is chosen for each $R$ value. The energies are measured from
the $^{40}$Ca+$\alpha$ threshold (dotted line). 
We can see a very large difference between these two curves, about 10 MeV
around $R = 1$ fm,
because of the contribution of the spin-orbit interaction in the solid line.
Here, as an effective nucleon-nucleon interaction,
 Volkov No.2 \cite{Vol} with Majorana exchange parameter of $M = 0.65$
has been adopted
for the central part, and G3RS \cite{G3RS}, which is a realistic
interaction determined to reproduce the nucleon-nucleon scattering phase shift, 
has been adopted for the spin-orbit part
(original strength of 600 MeV for the repulsive term and $-1050$ MeV
for the attractive term). 
Although the fine tuning of the interaction
is needed for the precise spectroscopy, at the moment we can say that
the effect of the spin-orbit interaction is quite large and that is clearly 
described by transforming $\alpha$ cluster states to the $jj$-coupling shell model one.

For $^{52}$Fe, the $jj$-coupling wave function is easily prepared
by placing our $^{12}$C (three quasi cluster) around the $^{40}$Ca core
introduced in the previous subsections. Here twelve nucleons of $^{12}$C
are excited to the $pf$ shell ($f_{7/2}$) due to the antisymmetrization effect.
The details for changing three $\alpha$ clusters into
three quasi clusters is shown in Ref.~\cite{Suhara},
and here we just add the $^{40}$Ca core introduced in the previous subsection in the center
of the three quasi clusters.

The expectation value of the one-body spin-orbit operator 
($\sum_i {\bf l_i} \cdot {\bf s_i}$)
for $^{52}$Fe
is shown in Fig.~\ref{Ti-ls} (dotted line).
The distance parameter for three quasi clusters are taken as $R = 0.25$ fm
($R$ in Eqs.~\eqref{ca-1}$\sim$\eqref{ca-10} for the $^{40}$Ca core part is set to 0.01 fm,
smaller than the value for the quasi cluster),
and $\Lambda$ is changed from 0 to 1.
The eigen value for one-body spin-orbit operator is 3/2
for a nucleon in the $f_{7/2}$ orbit,
and for $^{52}$Fe, the value should be 18 because of twelve nucleons in $f_{7/2}$.
In Fig.~\ref{Ti-ls} this is achieved at $\Lambda$ = 1.

\section{Summary}\label{summary}

We have shown that AQCM, which is a method 
to describe a transition from the $\alpha$-cluster wave function to the $jj$-coupling shell model
wave function, can be extended to heavier region.
In this model, this cluster-shell transition is characterized by only two parameters; 
$R$ representing the distance between $\alpha$ clusters 
and $\Lambda$ describing the breaking of $\alpha$ clusters,
and the contribution of the spin-orbit interaction, very important in the shell model,
can be taken into account starting with the $\alpha$ cluster model wave function. 
The $jj$-coupling shell model states are realized at the limit of $R \to 0$ and $\Lambda \to 1$.
We have shown the generality of this model in 
various $4N$ nuclei from $^4$He to $^{52}$Fe.

For $sd$ shell nuclei, we discussed for instance
in $^{28}$Si that there are two ways
to transform $\alpha$ cluster model wave function
to the closed $d_{5/2}$ configuration of the $jj$-coupling wave function,
which become identical at the limit of $R \to 0$ and $\Lambda \to 1$.
Furthermore,
this method can be extended to $pf$ shell nuclei.
The $\alpha$ cluster states have been widely discussed in $^{44}$Ti and $^{52}$Fe,
and smooth transition of the wave functions of these nuclei
from $\alpha$ cluster states to the lowest $jj$-coupling shell model configurations
can be straightforwardly described. 

As a future work, the description of particle-hole states is on going.
We have shown the transformation from the $\alpha$ cluster  wave function
to the shell model states, where the last nucleon occupy the spin-orbit favored orbit,
including the subshell closure configurations. 
Description for the 
excitation of one or two nucleon(s) from such spin-orbit favored orbits 
to unfavored orbits is now on going. 
This would be the last step to combine cluster and
shell model descriptions for the purpose of establishing an unified view of
the nuclear structure.

Also, this model could be applied to even heavier region
for the purpose of microscopic understanding of $\alpha$ decay.
The $\alpha$ cluster is formed in the surface region of the nucleus, but it should be broken inside
the surface. In general, shell models underestimate 
the decay probabilities 
(because of the small model space for the description of $\alpha$ formation
around the surface)
and cluster models overestimate
(because of the absence of $\alpha$ breaking components).
Here mixing the shell and cluster components is essential
to obtain the experimental width \cite{Varga}. 
Since our model gives 
a transparent view of smooth transition,
it could simplify the
quantitative description of the $\alpha$ decay.

\begin{acknowledgments}
Numerical calculation has been performed at Yukawa Institute for Theoretical Physics,
Kyoto University. This work was supported by JSPS KAKENHI Grant Numbers 
716143900002 (N.I.) and 15K17662 (T.S.).
\end{acknowledgments}

\end{document}